\documentclass[showpacs,preprintnumbers,amsmath,amssymb]{revtex4}

\usepackage{epsfig}
\usepackage{dcolumn}
\usepackage{bm}
\usepackage{citesort}
\date{\today}
\newcommand{\bmat}{\left(\begin{array}}

\newcommand{\emat}{\end{array}\right)}
\newcommand{\be}{\begin{equation}}
\newcommand{\ee}{\end{equation}}
\newcommand{\bea}{\begin{eqnarray}}
\newcommand{\eea}{\end{eqnarray}}

\def\lsim{\raise0.3ex\hbox{$\;<$\kern-0.75em\raise-1.1ex\hbox{$\sim\;$}}}
\def\gsim{\raise0.3ex\hbox{$\;>$\kern-0.75em\raise-1.1ex\hbox{$\sim\;$}}}

\def\OMIT#1{}

\newcommand{\nn}{\nonumber}

\begin{document}
\renewcommand{\thefootnote}{\fnsymbol{footnote}}

\baselineskip 16pt
\title{ $ B\to \rho K$ and  $
B\to \pi K^*$ decays in $SCET$}
\author{Gaber Faisel}
\affiliation{ Department of Physics and Center for Mathematics and Theoretical Physics,
National Central University, Chung-li, TAIWAN 32054.}

\affiliation{Egyptian Center for Theoretical Physics, Modern
University for Information and Technology, Cairo, Egypt}

\begin{abstract}

Exploring hints of New Physics in  the decay modes $B \to \pi K^*$
and $B\to\rho K$ can shed light on  the $B \to K\pi$ puzzle. In
this talk, we discuss supersymmetric contributions to the direct
CP asymmetries  of the decays $B \to \pi K^*$ and $B\to\rho K$
within Soft Collinear Effective Theory. We consider non-minimal
flavor SUSY contributions mediated by gluino exchange and apply
the  Mass Insertion Approximation (MIA) in the analysis. We show
that gluino contributions can enhance the CP asymmetries and
accommodate the experimental results.
\end{abstract}
\maketitle
\section{ Introduction}

Soft Collinear Effective Theory (SCET)  is an effective field
theory describing the dynamics of highly energetic particles
moving close to the light-cone interacting with a background field
of soft quanta\cite{Bauer:2000ew,Bauer:2000yr,Bauer:2002aj,
Chay:2003zp,Chay:2003ju,Bauer:2004tj,Jain:2007dy,Wang:2008rk,Fleming:2009fe,Faisel:2011kq}.
It provides a systematic and rigorous way to deal with the decays
of the heavy hadrons that involve different energy scales.
Moreover, the power counting in SCET helps to reduce the
complexity of the calculations and
 the factorization formula provided by SCET is
perturbative to all powers in $\alpha_s$ expansion.

 We can classify two different effective theories: SCET$_{I}$ and
SCET$_{II}$ according to the momenta modes in the process under
consideration. SCET$_{I}$  is applicable in the processes in which
the  momenta modes are the  collinear and the  ultra soft  as in
the inclusive decays of a heavy meson
 such as $B\rightarrow X^*_s\gamma$ at the end point region
 and $e^-p\rightarrow e^-X$  at the threshold region in which there
 are only collinear and ultra soft  momenta modes.
  SCET$_{II}$  is applicable to the semi-inclusive or exclusive decays of a heavy
 meson such as $B\rightarrow D\pi$, $B\rightarrow K\pi$, $B\rightarrow \pi\nu e$,....etc
 in which there  are only collinear and soft momenta modes.

The amplitude of $ B\to {M_1M_2}$ where $M_1$ and $M_2$ are light
mesons  in SCET can be written as follows \bea{\cal A}_{B\to
M_1M_2}^{SCET}
 &=& {\cal A}_ {B\to M_1M_2}^{LO} +  {\cal A}_{B\to M_1M_2}^ {\chi} +
 {\cal A}_{B\to M_1M_2}^{ann} +  {\cal A}_{B\to M_1M_2}^{c.c} \eea

Here ${\cal A}_ {B\to M_1M_2}^{LO}$ denotes the leading order
amplitude in the expansion $1/m_b$, $ {\cal A}_{B\to M_1M_2}^
{\chi}$ denotes the chirally enhanced penguin amplitude,  $ {\cal
A}_{B\to M_1M_2}^{ann}$ denotes the annihilation amplitude and $
{\cal A}_{B\to M_1M_2}^{c.c}$ denotes the long distance  charm
penguin contributions. In the following we give a brief account
for each amplitude.

At leading power in $(1/m_b)$ expansion, the full QCD effective
weak Hamiltonian of the $\Delta_B=1$ decays is matched into the
corresponding weak Hamiltonian in $SCET_I$ by integrating out  the
hard scale $ m_b $. Then, the $SCET_I$ weak Hamiltonian is
matched into the weak Hamiltonian $SCET_{II}$ by integrating out
the hard collinear modes with $p^2\sim \Lambda m_b$ and the
amplitude of the $\Delta_B=1$ decays at leading order in $\alpha_s$ expansion
can be obtained via~\cite{Bauer:2002aj}:

 \begin{eqnarray}
 {\cal A}^{LO}_{B\rightarrow M_1M_2}&=& - i\big\langle
M_1M_2 \big| H^{SCET_{II}}_W \big|
\bar{B}\big\rangle\nonumber\\&=&\frac{G_F
m_B^2}{\sqrt{2}}\Big(f_{M_1}[\int^1_0
 dudzT_{M_1J}(u,z)\zeta^{B M_2}_J(z)\phi_{M_1}(u)\nonumber\\
  &+& \zeta^{B M_2}\int^1_0
 du T_{M_1\zeta}(u)\phi_{M_1}(u)]+(M_1 \leftrightarrow
 M_2)\Big).\label{amp1}
 \end{eqnarray}

At leading order in $\alpha_s$ expansion, the parameters $\zeta^{B
(M_1,M_2)}$, $\zeta_J^{B (M_1,M_2)}$ are treated
 as hadronic parameters and can be determined through the $\chi^2$ fit method using  the
 non leptonic decay  experimental data of  the branching fractions and CP
 asymmetries. The hard kernels $T_{(M_1,M_2)\zeta}$ and $T_{(M_1,M_2) J}$ can be expressed in terms of
 $c_i^{(f)}$ and $b_i^{(f)}$ which are functions of the
 Wilson coefficients  as follows \cite{Jain:2007dy}
 \begin{eqnarray} \label{Tz}
 T_{1\zeta}(u) &=
  {\cal C}_{u_L}^{BM_2} \, {\cal C}_{f_Lu}^{M_1} \, c_1^{(f)}(u)
  +   {\cal C}_{f_L}^{BM_2} \, {\cal C}_{u_Lu}^{M_1} \, c_2^{(f)}(u)
  \nn \\
 & +   {\cal C}_{f_L}^{BM_2} \, {\cal C}_{u_Ru}^{M_1} \, c_3^{(f)}(u)
  +  {\cal C}_{q_L}^{BM_2} \, {\cal C}_{f_Lq}^{M_1} \, c_4^{(f)}(u)
  ,\nn\\
 T_{1 J}(u,z) &=
   {\cal C}_{u_L}^{BM_2} \, {\cal C}_{f_Lu}^{M_1} \, b_1^{(f)}(u,z)
  +   {\cal C}_{f_L}^{BM_2} \, {\cal C}_{u_Lu}^{M_1} \, b_2^{(f)}(u,z)
   \nn\\
 &+   {\cal C}_{f_L}^{BM_2} \, {\cal C}_{u_Ru}^{M_1} \, b_3^{(f)}(u,z)
  +  {\cal C}_{q_L}^{BM_2} \, {\cal C}_{f_Lq}^{M_1} \, b_4^{(f)}(u,z).
\end {eqnarray}
here $f$ stands for $d$ or $s$ and  ${\cal C}_i^{BM}$ and ${\cal
C}_{i}^M$ are Clebsch-Gordan coefficients that depend on the
flavor content of the final state mesons.  $c_i^{(f)}$ and
$b_i^{(f)}$ are given by \cite{Bauer:2004tj}
\begin{eqnarray}\label{cmatch}
c_{1,2}^{(f)}&=&\lambda_u^{(f)}\Big[C_{1,2}+\frac{1}{N}C_{2,1}\Big]
-\lambda_t^{(f)}\frac{3}{2}\Big[\frac{1}{N}C_{9,10}
+C_{10,9}\Big]+ \Delta c_{1,2}^{(f)},\nonumber\\
c_3^{(f)}&=&-\frac{3}{2}\lambda_t^{(f)}\Big[C_7+\frac{1}{N}C_8\Big]
+ \Delta c_3^{(f)},\nonumber\\
c_4{(f)}&=&-\lambda_t^{(f)}\Big[\frac{1}{N}C_3+C_4
-\frac{1}{2N}C_9-\frac{1}{2}C_10\Big]+ \Delta c_4^{(f)},
\end{eqnarray}
and
\begin{eqnarray}
b_{1,2}^{(f)}&=&\lambda_u^{(f)}\Big[C_{1,2}+\frac{1}{N}
\Big(1-\frac{m_b}{\omega_3}\Big)C_{2,1}\Big]-
\lambda_t^{(f)}\frac{3}{2}\Big[C_{10,9}+\frac{1}{N}
\Big(1-\frac{m_b}{\omega_3}\Big)C_{9,10}\Big]+ \Delta b_{1,2}^{(f)},\nonumber\\
b_3^{(f)}&=&-\lambda_t^{(f)}\frac{3}{2}\Big[C_7
+\Big(1-\frac{m_b}{\omega_2}\Big)\frac{1}{N}C_8\Big]+ \Delta b_3^{(f)},\nonumber\\
b_4^{(f)}&=&-\lambda_t^{(f)}\Big[C_4+\frac{1}{N}
\Big(1-\frac{m_b}{\omega_3}\Big)C_3\Big]+\lambda_t^{(f)}\frac{1}{2}
\Big[C_{10}+\frac{1}{N}\Big(1-\frac{m_b}{\omega_3}\Big)C_9\Big]+
\Delta b_4^{(f)},
\end{eqnarray}
where $\omega_2=m_bu$ and $\omega_3=-m_b\bar{u}$. $u$ and
$\bar{u}=1-u $ are momentum fractions for the quark and antiquark
$\bar{n}$ collinear fields. The $\Delta c_i^{(f)}$ and $\Delta
b_i^{(f)}$ denote terms depending on $\alpha_s$ generated by
matching from $H_W$. The ${\cal O}(\alpha_s)$ contribution to
$\Delta c_i^{(f)}$ has been calculated
in Refs.(\cite{Beneke:1999br,Beneke:2000ry,Chay:2003ju}) and later in Ref.
(\cite{Jain:2007dy}) while the ${\cal O}(\alpha_s)$contribution to
$\Delta b_i^{(f)}$ has been calculated
in Refs.(\cite{Beneke:2005vv,Beneke:2006mk,Jain:2007dy}).

Corrections of order $\alpha_s(\mu_h)(\mu_M\Lambda/m_b^2)$ where
$\mu_M$ is the chiral scale parameter generate the so called
Chirally enhanced penguins amplitude $ {\cal A}_{B\to M_1M_2}^
{\chi}$\cite{Jain:2007dy}. $\mu_M$ for kaons and pions can be of
order $(2GeV)$ and therefore chirally enhanced terms can compete
with the order $\alpha_s(\mu_h)(\Lambda/m_b)$ terms.   The
chirally enhanced amplitude for  $B\to M_1M_2$ decays is given
by\cite{Jain:2007dy}

\begin{eqnarray}\label{chienhanced} A^\chi(\bar B\to M_1 M_2) &=& \frac{G_F
m_B^2}{\sqrt2}  \bigg\{- \frac{\mu_{M_1}  f_{M_1}}{3m_B}
\zeta^{BM_2}  \int_0^1 du R_{1}(u)\phi_{pp}^{M_1}(u)
   + (1\leftrightarrow 2)\nonumber\\
   &-&  \frac{\mu_{M_1} f_{M_1}}{3m_B} \int_0^1 du dz
   R_{1}^{J}(u,z) \zeta_J^{BM_2}(z) \phi_{pp}^{M_1}(u)
 + (1\leftrightarrow 2) \nonumber\\ & -& \frac{\mu_{M_2}f_{M_1}}{6 m_B}\int_0^1 dudz
   R_{1}^{\chi}(u,z) \zeta_{\chi}^{BM_2}(z) \phi^{M_1}(u)
 +(1\leftrightarrow 2) \bigg\}
 \end{eqnarray}

The factors $\mu_M$ are generated by pseudoscalars and so they
vanish for vector mesons~\cite{Jain:2007dy}. The pseudoscalar
light cone amplitude $\phi_{pp}^M(u)$ is defined as
\cite{Hardmeier:2003ig,Arnesen:2006vb}
\begin{equation}
\phi_{pp}^P(u) =3u[\phi_p^P(u)+\phi_\sigma^{P\prime}(u)/6+ 2
f_{3P}/(f_P \mu_P) \int dy'/y' \phi_{3P}(y-y',y)].
 \end{equation}

 The hard kernels $R_{K},R_{\pi},R_{K}^J,R_{\pi}^J,R_{K}^{\chi}$
and $ R_{\pi}^{\chi}$ can be expressed in terms of Clebsch-Gordan
coefficients for the different final states as\cite{Jain:2007dy}
\begin{eqnarray}
\label{chihard} R_{1}(u) &=  {\cal C}_{q_R}^{BM_{2}} {\cal
C}_{f_Lq}^{M_{1}}
  \Big[ c^{\chi}_{1(qfq)}+\frac{3}{2}e_q \,c^{\chi}_{2(qfq)} \Big]
\,, \\
R_{1}^J(u,z) &=  {\cal C}_{q_R}^{BM_{2}} {\cal C}_{f_Lq}^{M_{1}}
  \Big[ b^{\chi}_{3(qfq)}+\frac{3}{2}e_q \,b^{\chi}_{4(qfq)} \Big]
\,, \nn \\
R_{1}^{\chi}(u,z) &=
  {\cal C}_{q_L}^{BM_{2}} {\cal C}_{f_Lq}^{M_{1}} \, b^{\chi}_{1(qfq)}
 + {\cal C}_{u_L}^{BM_{2}} {\cal C}_{f_Lu}^{M_{1}}\,  b^{\chi}_{1(ufu)} \nn \\
& +{\cal C}_{f_L}^{BM_{2}} {\cal C}_{u_L u}^{M_{1}}\,
b^{\chi}_{1(fuu)}
  +{\cal C}_{f_L}^{BM_{2}} {\cal C}_{u_R u}^{M_{1}}\, b^{\chi}_{2(fuu)} \, .\nn
\end{eqnarray}
Summation over $q=u,d,s$ is implicit and  $c_{i}^{\chi}$ and
$b_{i}^{\chi}$ are expressed in terms of the short-distance Wilson
coefficients and can be found in Ref.\cite{Jain:2007dy}.

 Annihilation amplitudes ${\cal A}_{B\to M_1M_2}^{ann}$ have been studied
 in PQCD and QCD factorization in Refs.(\cite{Keum:2000ph,Lu:2000em,Beneke:2001ev,Kagan:2004uw}).
Within SCET, the annihilation contribution becomes factorizable
and  real at leading order,  $ {\cal O}(\alpha_s(m_b)\Lambda/m_b)$
\cite{Manohar:2006nz}.  Their size are small and contains large
uncertainty compared to the other contributions\cite{Arnesen:2006vb,Jain:2007dy}.

  In SCET, charm penguins are  treated as non perturbative and its amplitude is parameterized as
 \be {\cal A}_{B\to M_1 M_2}^{c.c}=|{\cal A}_{B\to M_1
M_2}^{c.c}|e^{i \delta_{cc}} \ee where $\delta_{cc}$ is the strong
phase of the charm penguin. The modulus and the phase of the charm penguin
are fixed through the fitting with non leptonic decays in a
similar way to the hadronic parameters $\zeta^{B (M_1,M_2)}$,
$\zeta_J^{B (M_1,M_2)}$.

\section{SM contribution to the CP asymmetries of $B\to \pi K^*$ and $B\to
\rho K$ decays}\label{sec:SM}

Withen SCET, the predicted  branching ratios of the decay modes
$B\to \pi K^*$ and $B\to \rho K$  are in agreements with their
corresponding experimental values in most of the decay
modes\cite{Wang:2008rk,Faisel:2011qt}. On the other hand, the SM
predictions for the CP asymmetries of $B^+\to \pi^0 K^{*\,+}$ has
different sign in comparison with the experimental measurement and
the predicted CP asymmetries in many of the decay modes are in
agreement with the experimental measurements due to the large
errors in these measurements\cite{Faisel:2011qt}. Moreover, the
predicted CP asymmetry of $\bar{B}\to \pi^0 \bar{K}^{*\,0}$ and
$B^+\to\rho^0 K^+$ disagree with the experimental results within
$1\sigma.$ error of the experimental data.  Note, SCET provides
large strong  phases and thus with new sources of weak CP
violation one can expect enhancement of these asymmetries. This
possibility will be discussed in the next section considering
supersymmetry (SUSY) as a possible candidate of physics beyond SM
that has new sources of weak phases.

\section{SUSY contributions to the CP asymmetries of$ B\to \rho K$ and $ B\to \pi K^*$}\label{sec:SUSY}

Supersymmetry  has new sources for CP violation which can account
for the baryon number asymmetry and affect other CP violating
observables in the B and K decays. The effects of these phases on
the CP asymmetries in semi-leptonic $\tau$ decays has been studied
in Refs.(\cite{Delepine:2006fv,Delepine:2007qg,Delepine:2008zzb}).
In SUSY, Flavor Changing Neutral Current(FCNC) and CP quantities
are sensitive to particular entries in the mass matrices of the
scalar fermions. Thus it is useful to adopt a model independent-
parametrization, the so-called Mass Insertion Approximation (MIA)
where all the couplings of fermions and sfermions to neutral
gauginos are flavor diagonal~\cite{Hall:1985dx}.  The Flavor
Changing structure of the $A-B$ sfermion propagator  is exhibited
by its non-diagonality and it can be expanded as
\begin{equation}
\langle \tilde{f}_{A}^{a}\tilde{f}_{B}^{b\ast }\rangle \simeq \frac{i\delta _{ab}}{k^{2}-\tilde{m}%
^{2}}+\frac{i(\Delta
_{AB}^{f})_{ab}}{(k^{2}-\tilde{m}^{2})^{2}}+O(\Delta ^{2}),
\end{equation}%
where $a,b=(1,2,3)$ are flavor indices, $\Delta$ are the
off-diagonal terms in the $(M^2_{\tilde{f}})_{AB}$ and $I$ is the
unit matrix. It is convenient to define a dimensionless quantity
$(\delta
_{AB}^{f})_{ab}\equiv (\Delta _{AB}^{f})_{ab}/\tilde{m}^{2}.$ As long as $%
(\Delta _{AB}^{f})_{ab}$ is smaller than $\tilde{m}^{2}$ we can
consider only the first order term in $(\delta _{AB}^{f})_{ab}$ of
the sfermion propagator expansion.

 The parameters $(\delta _{AB}^{f})_{ab}$ can be constrained
through vacuum stability argument~\cite{Casas:1996de},
experimental measurements concerning FCNC and CP violating
phenomena~\cite{Gabbiani:1996hi}. Recent studies about other
possible constraints can be found in
Refs.(~\cite{Crivellin:2008mq,Crivellin:2009ar,Crivellin:2010gw}).

The mass insertions $(\delta^u_{RL})_{32}$ and
$(\delta^u_{LR})_{32}$ are not constrained by $b \to s \gamma$
 and so we can set them as $(\delta^u_{RL})_{32}=(\delta^u_{LR})_{32}= e^{i\delta_u}$
 where $\delta_u$ is the phase that can vary from $-\pi$ to $\pi$.  It should be noted that in order
 to have a well defined Mass Insertion Approximation scheme, it is necessary to have
 $|(\delta^f_{AB})_{ab} | < 1$  but here in order
 to maximize the SUSY CP-violating contributions we take it of order one.
 Applying $b \to s \gamma$ constraints leads to the following parametrization~\cite{Huitu:2009st}

 \begin{equation}
 (\delta^d_{LL})_{23}= e^{i\delta_d}~~~~~~~~~~~ (\delta^d_{LR})_{23}=(\delta^d_{RL})_{23}= 0.01 e^{i\delta}\label{masinscons}
 \end{equation}

\begin{figure}[tbhp]
\includegraphics[width=6.5cm,height=7cm]{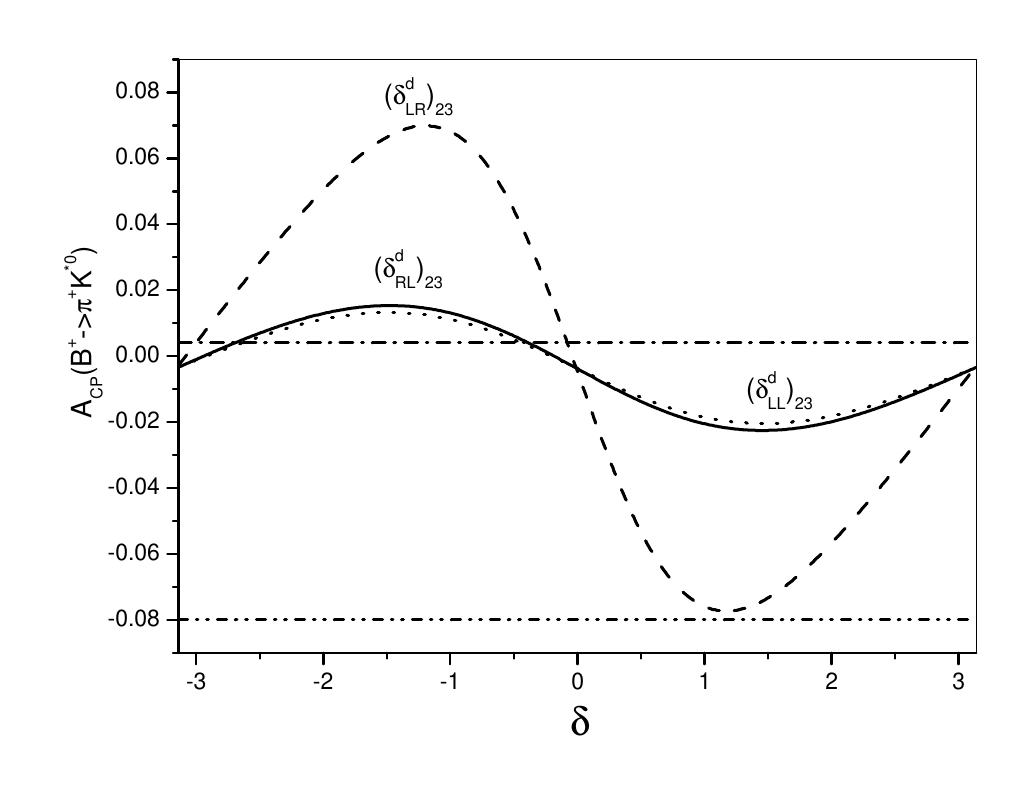}
\hspace{1.cm}
\includegraphics*[width=6.5cm,height=7cm]{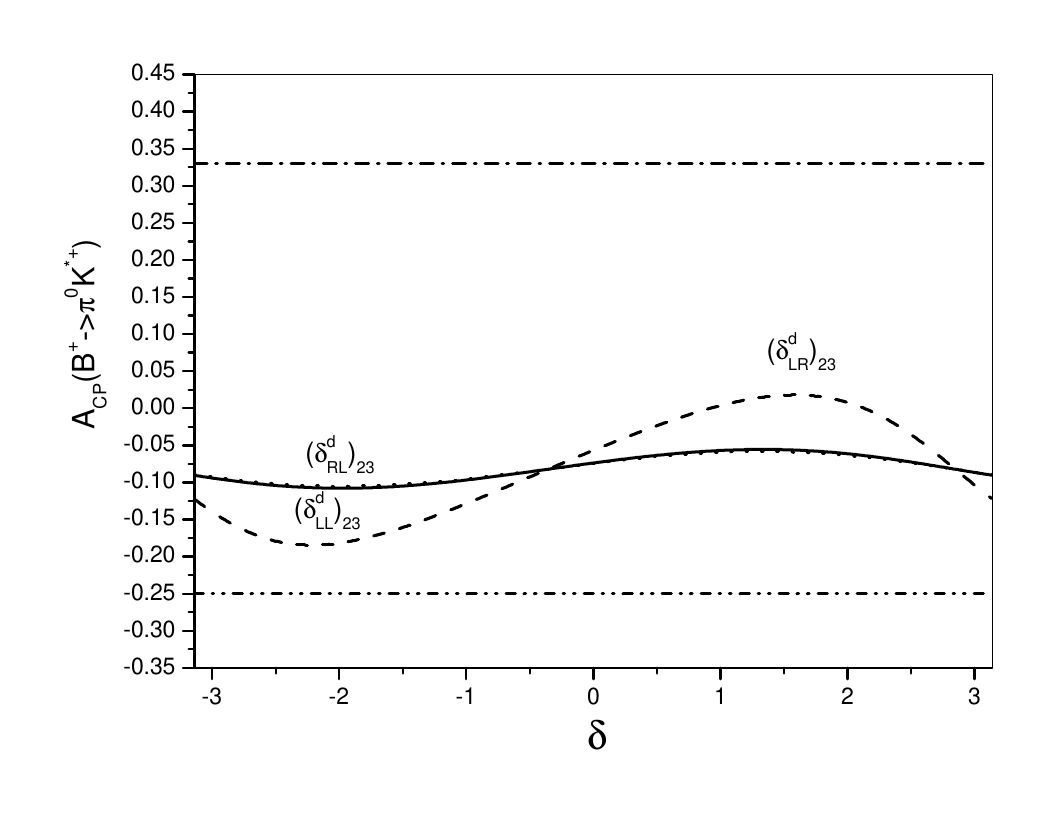}
\medskip
\caption{ CP asymmetries  versus the phase of the
$(\delta^{d}_{AB})_{23}$ where A and B denote the chirality i.e.
L, R. for 3 different mass insertions.
 The left diagram corresponds to  $A_{CP}(B^+\to \pi^+\bar{K}^{*\,0})$ while the right diagram corresponds to
 $A_{CP}(B^+\to \pi^0 K^{*\,+})$. In both diagrams we take only one mass insertion
per time and vary the phase of  from $-\pi$ to $\pi$. The
horizontal lines in both diagrams represent the experimental
measurement to $1\sigma$\cite{Faisel:2011qt}.} \label{singlemas}
\end{figure}

\begin{figure}[tbhp]
\includegraphics[width=6.5cm,height=7cm]{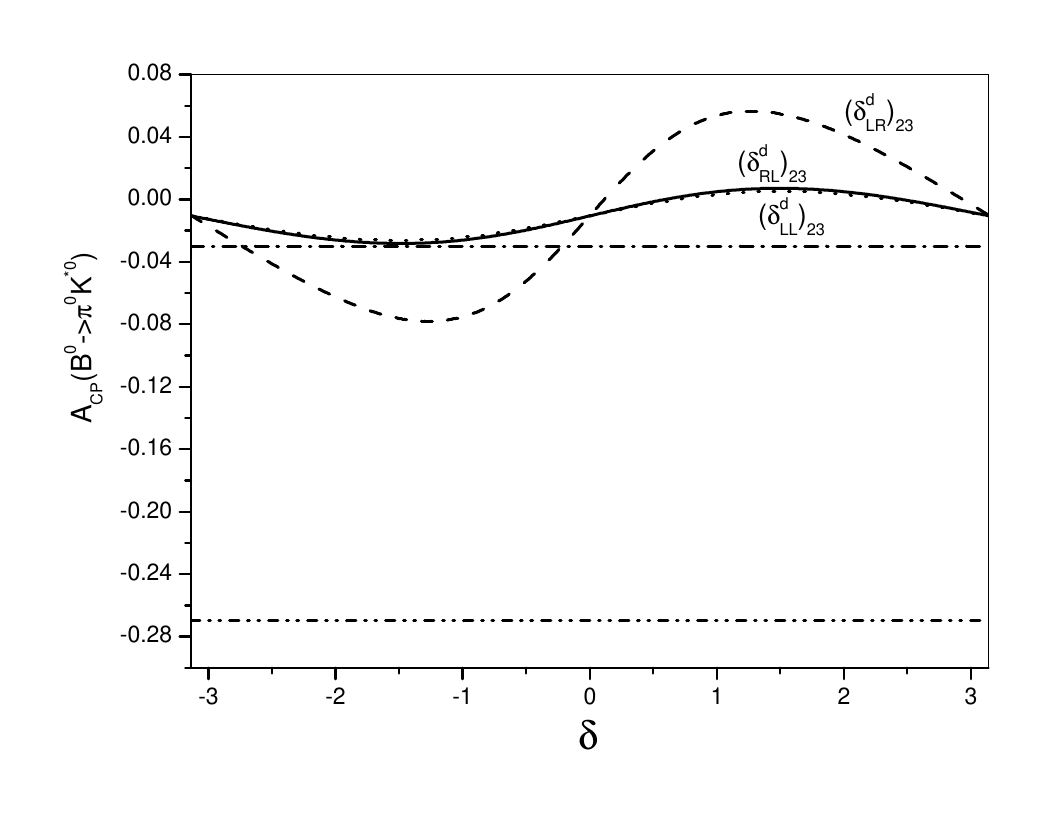}
\hspace{1.cm}
\includegraphics*[width=6.5cm,height=7cm]{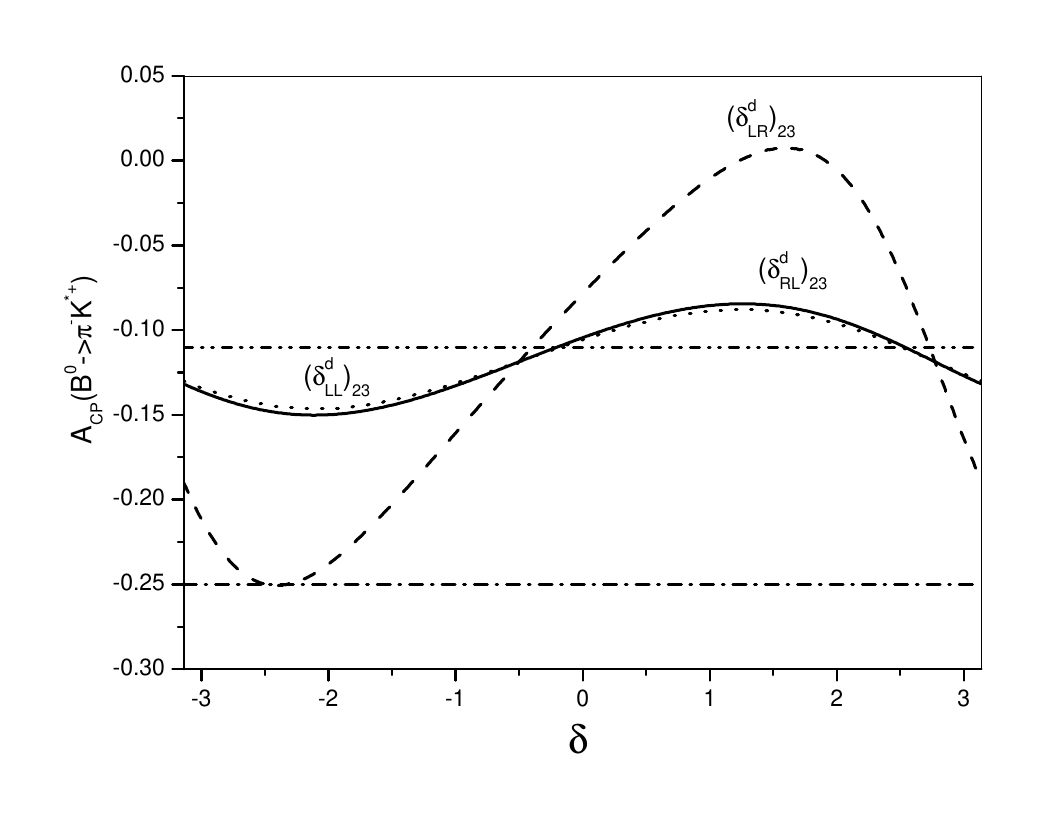}
\medskip
\caption{CP asymmetries  versus the phase of the
$(\delta^{d}_{AB})_{23}$ where A and B denote the chirality i.e.
L, R. for 3 different mass insertions.
 The left diagram corresponds to  $A_{CP}(B^0\to \pi^0\bar{K}^{*\,0})$ while the right diagram corresponds to
$A_{CP}(B^0\to \pi^-K^{*\,+})$. In both diagrams we take only one
mass insertion per time and vary the phase of  from $-\pi$ to
$\pi$. The horizontal lines in both diagrams represent the
experimental measurement to $1\sigma$\cite{Faisel:2011qt}.}
\label{singlemas2}
\end{figure}

\begin{figure}[tbhp]
\includegraphics[width=6.5cm,height=7cm]{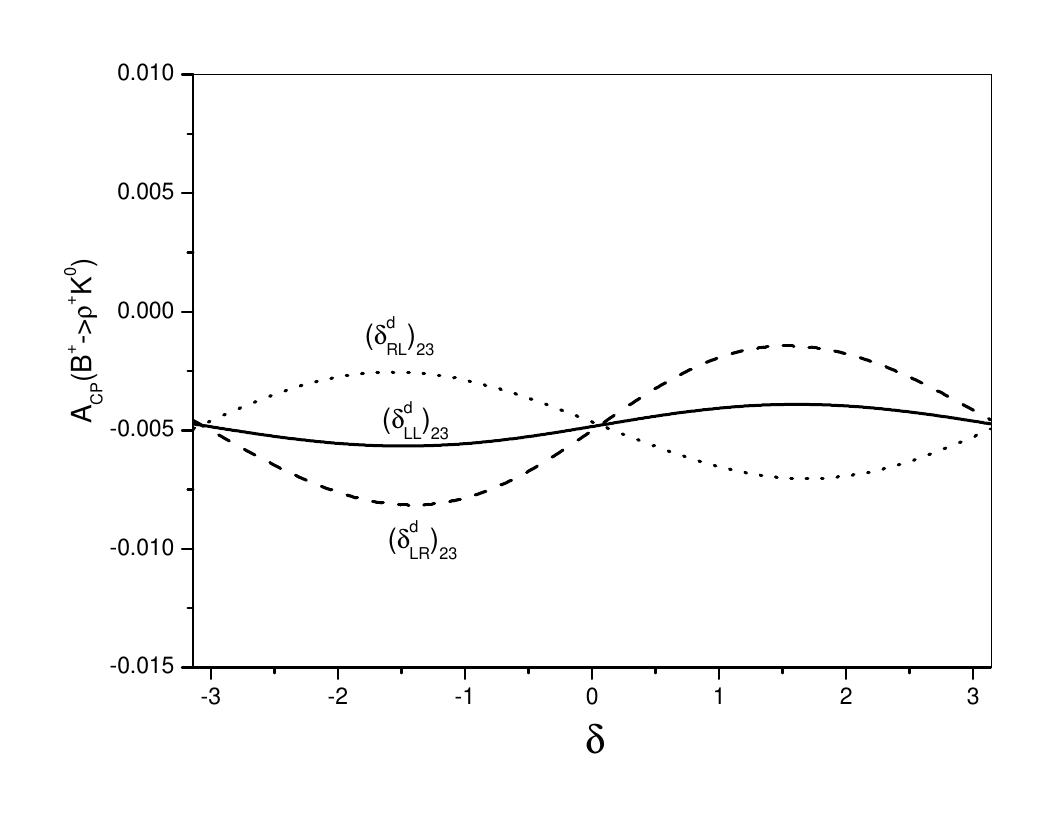}
\hspace{1.cm}
\includegraphics*[width=6.5cm,height=7cm]{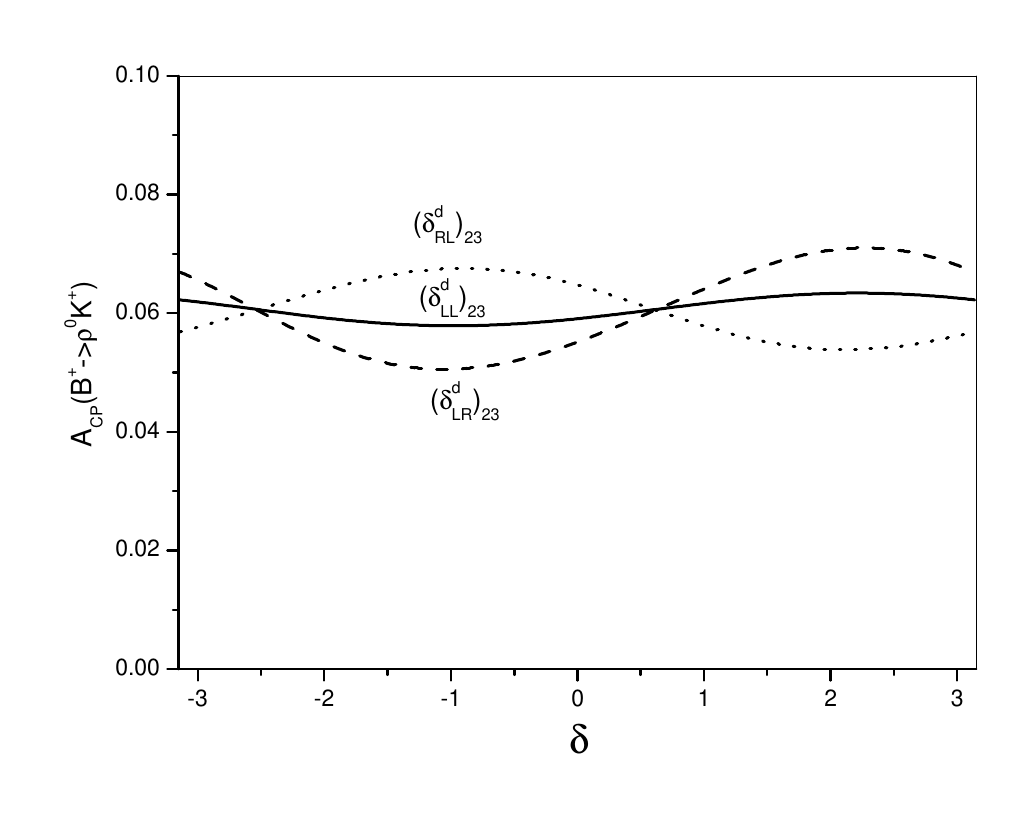}
\medskip
\caption{CP asymmetries  versus the phase of the
$(\delta^{d}_{AB})_{23}$ where A and B denote the chirality i.e.
L, R. for 3 different mass insertions.
 The left diagram corresponds to  $A_{CP}(B^+\to \rho^+ K^{0})$ while the right diagram corresponds to
$A_{CP}(B^+\to \rho^0 K^{+})$. In both diagrams we take only one
mass insertion per time and vary the phase of  from $-\pi$ to
$\pi$\cite{Faisel:2011qt}.} \label{singlemas3}
\end{figure}

\begin{figure}[tbhp]
\includegraphics[width=6.5cm,height=7cm]{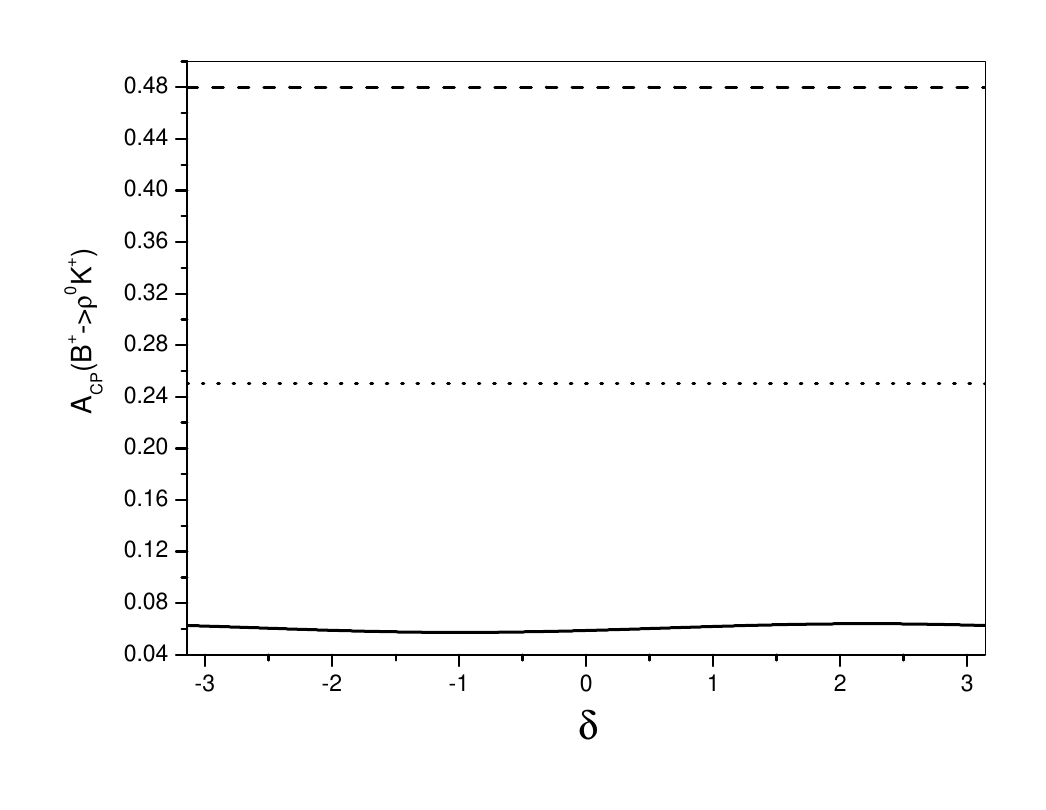}
\hspace{1.cm}
\includegraphics*[width=6.5cm,height=7cm]{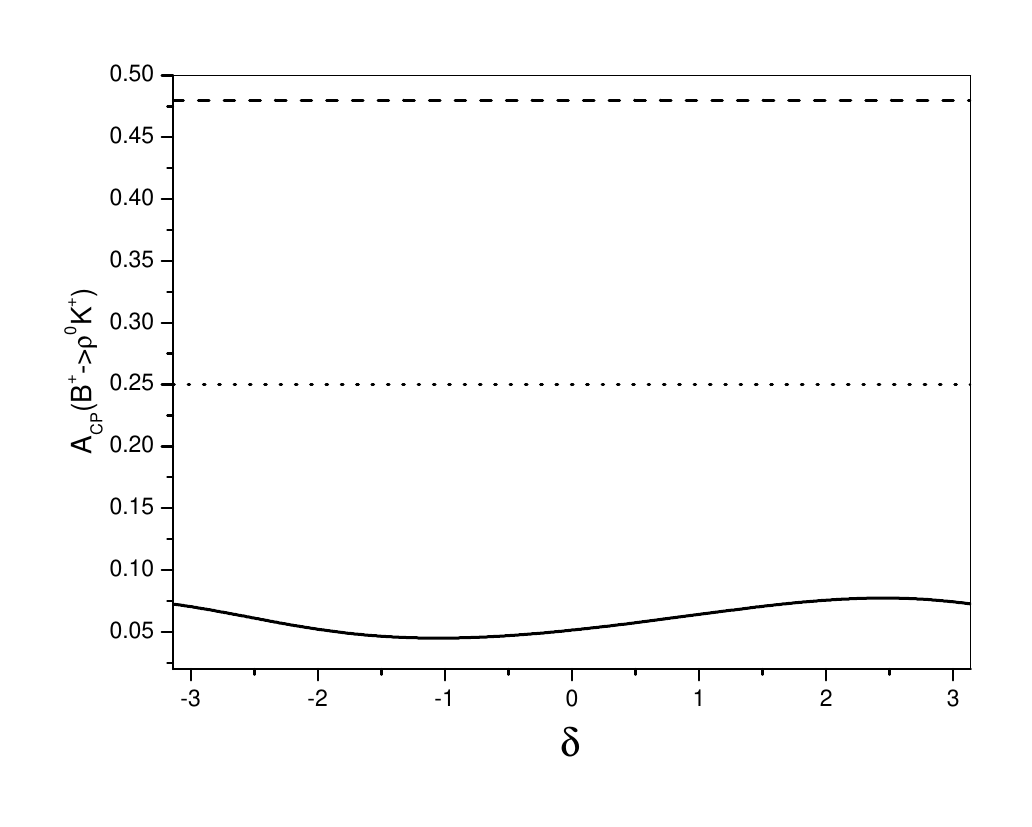}
\medskip
\caption{CP asymmetry of $A_{CP}(B^+\to \rho^0 K^{+})$ versus the
phase of the mass insertion  for 2 different mass insertions.
 The left diagram correspond to gluino contributions where we keep
 the two mass insertions $(\delta^{d}_{LR})_{23}$ and $(\delta^{d}_{RL})_{23}$
 and set the other mass insertions to zero.   The right diagram correspond to
both gluino and chargino contributions where we keep
 the two mass insertions $(\delta^{d}_{LR})_{23}$ and $(\delta^{u}_{RL})_{32}$
 and set the other mass insertions to zero.  The horizontal lines in both diagrams represent the experimental
measurements to $1\sigma$\cite{Faisel:2011qt}.} \label{singlemas4}
\end{figure}

 We consider two scenarios, the first one with a
single mass insertion where we keep only one mass insertion per
time and take the other mass insertions to be zero and the second
scenario with two mass insertions will be considered only in the
cases when one single mass insertion is not sufficient to
accommodate the experimental measurement. After setting the
different  mass insertions as mentioned above, we find that, the
terms that contain the mass insertions $(\delta^{u}_{RL})_{32}$
and $(\delta^{u}_{LR})_{32}$ will be small in comparison  with the
other terms and thus we expect that their contributions to the
asymmetries will be small. These terms are obtained from diagrams
mediated by the chargino exchange and thus we see that gluino
contributions give the dominant contributions. The results  are
presented in Figures(\ref{singlemas},\ref{singlemas2},\ref{singlemas3},
\ref{singlemas4})

 In Figure\ref{singlemas},we show the  CP asymmetries,
$A_{CP}(B^+\to \pi^+\bar{K}^{*\,0})$ and  $A_{CP}(B^+\to \pi^0
K^{*\,+})$  versus the phase of the $(\delta^{d}_{AB})_{32}$
 where A and B denote the chirality i.e. L and R. for 3 different mass insertions. The horizontal lines
in both diagrams represent the experimental measurements to
$1\sigma$.  As can be seen from  Figure\ref{singlemas}~left, for
all gluino mass insertions, the value of the CP asymmetry
$A_{CP}(B^+\to \pi^+\bar{K}^{*\,0})$ is enhanced to accommodate
the experimental measurement of the asymmetry within $1\sigma$ for
many values of the phase of the mass insertions. On the other
hand,Figure \ref{singlemas}right shows that  the CP asymmetry
$A_{CP}(B^+\to \pi^0 K^{*\,+})$ is enhanced  to accommodate  the
experimental measurement within $1\sigma$ for all values of the
phase of the mass insertions.

In Fig.\ref{singlemas2}  we show  the two asymmetries,
$A_{CP}(B^0\to \pi^0\bar{K}^{*\,0})$ and  $A_{CP}(B^0\to
\pi^-K^{*\,+})$ versus the phase of the $(\delta^{d}_{AB})_{32}$
as before.  Clearly from  Fig.\ref{singlemas2} left,
$A_{CP}(B^0\to \pi^0\bar{K}^{*\,0})$  lies within $1\sigma$ range
of its experimental value for  many values of the phase of the
mass insertion $(\delta^{d}_{LR})_{23}$ only. The reason is that
the two mass insertions $(\delta^d_{LL})_{23}$ and
$(\delta^d_{RL})_{23}$ have equal contributions to the CP
asymmetries which  will be smaller than the case of using
$(\delta^d_{LR})_{23}$.  On the other hand, Fig.\ref{singlemas2}
right, one sees that  $A_{CP}(B^0\to \pi^-K^{*\,+})$ can be
accommodated within $1\sigma$ for many values of the phase of the
three gluino mass insertions.

Finally we present  the CP asymmetries of the decay modes $B^+\to
\rho^+ K^{0}$ and $B^+\to \rho^0 K^{+}$ in Fig.(\ref{singlemas3}).
In Fig.(\ref{singlemas3}) we do not show the horizontal lines
representing the $1\sigma$ range of the experimental measurement
as the three curves of the $A_{CP}(B^+\to \rho^+ K^{0})$
corresponding to the three gluino mass insertions totally lie in
this $1\sigma$ range for all values of the phase of the mass
insertions.  On the other hand, Fig.\ref{singlemas3} right, we see
that $A_{CP}(B^+\to \rho^0 K^{+})$  can not be accommodated within
$1\sigma$ for any value of the phase of all gluino mass
insertions. This motivates us to consider the second scenario with
two mass insertions which is shown in Fig.\ref{singlemas4}.

\begin{figure}[tbhp]
\includegraphics[width=6.5cm,height=7cm]{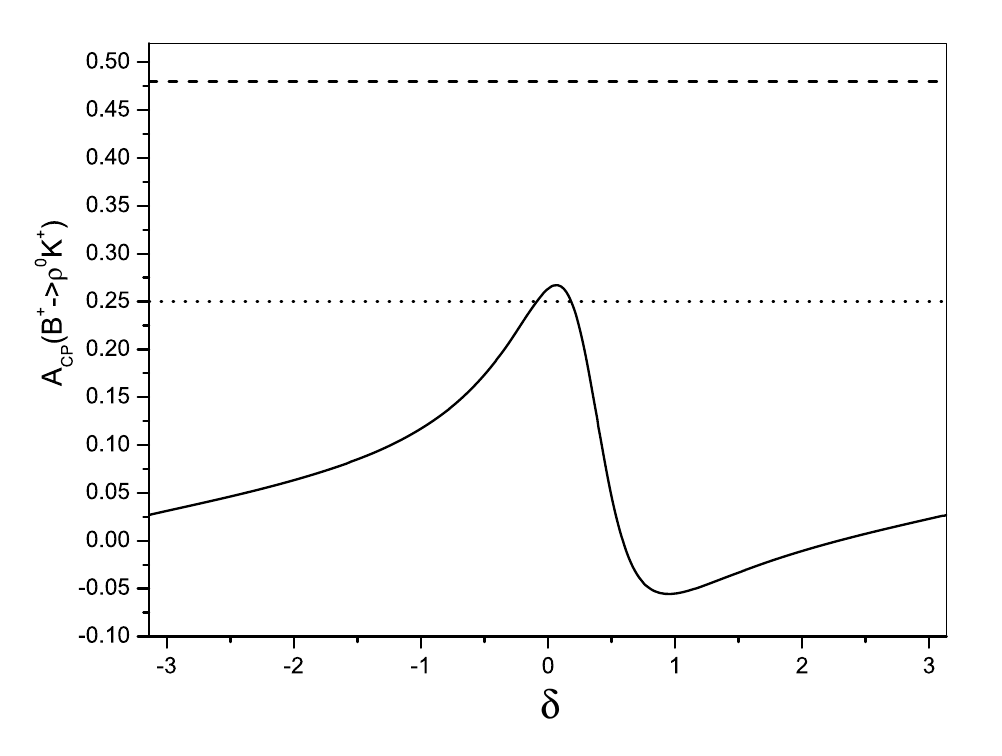}
\medskip
\caption{CP asymmetry of $A_{CP}(B^+\to \rho^0 K^{+})$ versus the
phase of the mass insertion  for 2 different mass insertions
correspond to gluino contributions where we keep the two mass
insertions $(\delta^{d}_{LR})_{23}$ and $(\delta^{d}_{LL})_{23}$
and set the other mass insertions to zero.  We assume that the two
mass insertion have equal phases and we vary the phase  from
$-\pi$ to $\pi$. The horizontal lines in the diagram represent the
experimental measurements to $1\sigma$\cite{Faisel:2011qt}.}
\label{singlemas5}
\end{figure}

The left diagram correspond to gluino contributions where we keep
the two mass insertions $(\delta^{d}_{LR})_{23}$ and
$(\delta^{d}_{RL})_{23}$  and set the other mass insertions to
zero.  The right diagram correspond to both gluino and chargino
contributions where we keep  the two mass insertions
$(\delta^{d}_{LR})_{23}$ and $(\delta^{u}_{RL})_{32}$ and set the
other mass insertions to zero. In both diagrams we assume that the
two mass insertion have equal phases and we vary the phase  from
$-\pi$ to $\pi$.  As can be seen from Fig.\ref{singlemas4} left,
two gluino mass insertions can not accommodate the experimental
measurement for any value of the phase of the mass insertion. On
the other hand from Fig.\ref{singlemas4} right, two mass
insertions one corresponding to chargino contribution and the
other corresponding to  gluino contribution can not  accommodate
the experimental measurements. We find that in order to
accommodate the CP symmetry in this case the Wilson coefficient
$C^{\tilde{g}}_{9}$ should be increased at least by a factor
$-6\pi/\alpha$ without violating any constraints on the SUSY
parameter space which is shown in Fig.\ref{singlemas5}.

\section{Conclusion}\label{sec:conclusion}

 In this talk we discussed SUSY contributions to the direct CP asymmetries of  $B \to \pi K^*$ and
$B\to\rho K$ decays within Soft Collinear Effective Theory. We
considered non minimal flavor  SUSY models and applied  the mass
insertion approximation to analyze SUSY contributions to the CP
asymmetries of $B \to \pi K^*$ and $B\to\rho K$ decays.  We  show
that in most decay channels, direct CP asymmetries can be
significantly enhanced by the SUSY contributions mediated by
gluino exchange and thus accommodate the experimental results. For
the decay mode $B^+\to \rho^+ K^{0}$, we find that the enhancement
is not enough to accommodate the CP asymmetry.  To accommodate the
CP asymmetry of $B^+\to \rho^+ K^{0}$, the Wilson coefficient
$C^{\tilde{g}}_{9}$ should be increased at least by a factor
$-6\pi/\alpha$.

\section*{Acknowledgement}

Gaber Faisel would like to thank David Delepine and Mostafa
Shalaby for their collaborations in the realization of this work.
Gaber Faisel's  work is supported by the National Science Council
of R.O.C. under grants NSC 99-2112-M-008-003-MY3 and NSC
99-2811-M-008-085.


\end{document}